\documentclass[preprint2]{aastex631}

\usepackage{graphicx}
\usepackage{natbib}
\usepackage{booktabs}
\usepackage{amsmath, physics}
\usepackage{enumitem}

\newcommand{\Cornell}{Cornell Center for Astrophysics and Planetary Science, and Department of Astronomy, Cornell University, Ithaca, NY 14853, USA}
\newcommand{\Columbia}{Department of Astronomy, Columbia University, 550 West 120th Street, New York, NY 10027, USA}
\newcommand{\KZA}{University of Malta, Institute of Space Sciences and Astronomy, Malta}
\newcommand{\NIJ}{Department of Astrophysics/IMAPP, Radboud University, Nijmegen, The Netherlands}
\newcommand{\SETI}{SETI Institute, Mountain View, CA 94043, USA}
\newcommand{\UCB}{Breakthrough Listen, University of California, Berkeley, CA 94720, USA}
\newcommand{\UToronto}{Dunlap Institute for Astronomy \& Astrophysics, University of Toronto, 50 St.~George Street, Toronto, ON M5S 3H4, Canada}

\shorttitle{Galactic Center Pulsar Searches}
\shortauthors{Suresh et al.}
\begin{document}

\title{4--8~GHz Fourier-domain Searches for Galactic Center Pulsars}

\correspondingauthor{Akshay Suresh}
\email{as3655@cornell.edu}

\author[0000-0002-5389-7806]{Akshay Suresh}
\affiliation{\Cornell}
\affiliation{\UCB}

\author[0000-0002-4049-1882]{James M. Cordes}
\affiliation{\Cornell}

\author[0000-0002-2878-1502]{Shami Chatterjee}
\affiliation{\Cornell}

\author[0000-0002-8604-106X]{Vishal Gajjar}
\affiliation{\UCB}

\author[0000-0002-6341-4548]{Karen I. Perez}
\affiliation{\Columbia}

\author[0000-0003-2828-7720]{Andrew P. V. Siemion}
\affiliation{\UCB}
\affiliation{\NIJ}
\affiliation{\SETI}
\affiliation{\KZA}

\author[0000-0002-7042-7566]{Matt Lebofsky}
\affiliation{\UCB}

\author{David H. E. MacMahon}
\affiliation{\UCB}

\author[0000-0002-3616-5160]{Cherry Ng}
\affiliation{\UToronto}
\affiliation{\UCB}
\affiliation{\SETI}
%% Keep to less than 250 words.
\begin{abstract}
The Galactic Center (GC), with its high density of massive stars, is a promising target for radio transient searches. In particular, the discovery and timing of a pulsar orbiting the central supermassive black hole (SMBH) of our Galaxy will enable stringent strong-field tests of gravity and accurate measurements of SMBH properties. We performed multi-epoch 4--8~GHz observations of the inner $\approx 15$~pc of our Galaxy using the Robert C.~Byrd Green Bank Telescope in 2019 August--September. Our investigations constitute the most sensitive 4--8~GHz GC pulsar survey conducted to date, reaching down to a 6.1~GHz pseudo-luminosity threshold of $\approx$ 1~mJy~kpc$^2$ for a pulse duty cycle of 2.5$\%$. 
We searched our data in the Fourier-domain for periodic signals incorporating a constant or linearly changing line-of-sight  pulsar acceleration. We report the successful detection of the GC magnetar PSR~J1745$-$2900 in our data. Our pulsar searches yielded a non-detection of novel periodic astrophysical emissions above a $6\sigma$ detection threshold in harmonic-summed power spectra. We reconcile our non-detection of GC pulsars with inadequate sensitivity to a likely GC pulsar population dominated by millisecond pulsars. Alternatively, close encounters with compact objects in the dense GC environment may scatter pulsars away from the GC. The dense central interstellar medium may also favorably produce magnetars over pulsars.
\end{abstract}

% SECTION 1: INTRODUCTION
\section{Introduction} \label{sec:intro}
The central parsec of our Galaxy hosts a dense nuclear star cluster (NSC, \citealt{Schodel2007}) surrounding the supermassive black hole (SMBH), Sgr~A* of mass $M_{\rm SgrA*} \approx (4.30 \pm 0.01) \times 10^6 \ M_{\odot}$ \citep{GRAVITY2021}. The Galactic NSC, while primarily containing old, late-type stars ($\geq 10$~Gyr, \citealt{Schodel2020}), is home to a large population of neutron star and black hole (BH) progenitors, including young, massive main-sequence stars \citep{Ghez2005,Genzel2010} and Wolf-Rayet stars \citep{Paumard2001}. Recent detections of numerous X-ray binaries \citep{Hailey2018,Zhu2018} and compact steep-spectrum radio sources \citep{Chiti2016,Hyman2005,Hyman2009,Hyman2021,Zhao2020,Zhao2022} further indicate a likely abundance of neutron stars and stellar-mass BHs in the NSC. Considering multiwavelength constraints on the known neutron star population, \citet{Wharton2012} argued for the existence of $\sim 10^3$ radio pulsars actively beaming towards the Earth from the inner parsec of our Galaxy. Several of these pulsars may potentially reside in binaries, analogous to the profusion of millisecond pulsars\footnote{We define MSPs as pulsars with barycentric rotational periods, $P_0 < 30$~ms.} (MSPs) seen in globular clusters \citep{Ransom2008}. Additionally, a substantial MSP population \citep{Brandt2015,Lee2015,Bartels2016,Fragione2018} in the NSC has been postulated as a plausible explanation for the observed diffuse $\gamma$-ray excess around Sgr~A* \citep{Ackermann2014,Ajello2016}; another being dark matter annihilation \citep{Abazajian2014,Calore2015} at the Galactic Center (GC).  \\

Enabling powerful strong-field tests of gravity \citep{Wex1999,Kramer2004, Liu2012, Wex2014,Psaltis2016}, the discovery and timing of even a canonical pulsar\footnote{We define CPs as pulsars with $P_0 \geq 30$~ms.} (CP) in a binary with a BH will allow rigorous tests of the Cosmic Censorship Conjecture \citep{Penrose1969,Penrose1999} and the BH No Hair theorem. Furthermore, if orbiting close enough (binary orbital period, $P_b \lesssim 1$~yr) to Sgr~A*, regular pulsar timing efforts will permit accurate measurements of SMBH mass (anticipated precision $\simeq$ 1--10~$M_{\odot}$ with weekly observing cadence over five years, \citealt{Liu2012}), spin, and quadrupole moment. Finally, radio observations of  pulse dispersion, scattering, and Faraday rotation will provide unique probes of the turbulent, magneto-ionic central interstellar medium (ISM) of our Galaxy. \\

Motivated by the rich rewards of GC pulsar timing, numerous extensive surveys of the GC have been previously undertaken over a broad range of radio frequencies (e.g., \citealt{Johnston2006,Deneva2009,Macquart2010,Bates2011,Siemion2013,Eatough2013a,Eatough2013b,Eatough2021,Liu2021,Torne2021}, etc.). To date, these endeavors have revealed a single magnetar, namely PSR~J1745$-$2900 \citep{Eatough2013b} at $2\farcs4$ offset from Sgr~A*, and five pulsars, all located $\gtrsim 10'$ away from Sgr~A*. The hitherto non-detection of pulsars within a $10'$ radius of Sgr~A*, termed the ``missing pulsar problem,'' is often attributed to hyperstrong interstellar scattering in the direction of the GC \citep{Cordes1997,Lazio1998a,Lazio1998b,Cordes2002}. While pulse broadening measurements of PSR~J1745$-$2900 suggest otherwise \citep{Spitler2014}, it is unclear if a single line of sight towards the GC magnetar is representative of a possibly complex scattering structure at the GC \citep{Cordes2002,Schnitzeler2016,Dexter2017}. Alternatively, an abundance of highly magnetized massive stars in the NSC may preferentially produce magnetars over spin-driven pulsars \citep{Dexter2014}. \\

Aside from scattering, additional obstacles to GC pulsar discovery include the Galactic background temperature ($T_{\rm GC}(\nu)$: \citealt{Law2008}), free-free absorption by the ionized central ISM, and orbital motion (if in multi-object systems). While observations at high radio frequencies ($\nu \gtrsim 10$~GHz) help mitigate $T_{\rm GC} (\nu)$ and free-free absorption, pulsar emission also significantly weakens with increasing $\nu$ (period-averaged flux density, $S_{\nu} \propto \nu^{-1.4\pm 1.0}$, \citealt{Bates2013}). Weighing various chromatic challenges to GC pulsar discovery, \citet{Rajwade2017} recommended 9--13~GHz as the optimal observing band for GC pulsar surveys. However, the observed broad spread of pulsar spectral indices \citep{Bates2013} mandates continued broadband monitoring to detect elusive GC pulsars. \\

Orbital motion limits pulsar detection by smearing harmonics of the pulsar rotational frequency ($f_0 = 1/P_0$) in the power spectrum. Standard Fourier-domain algorithms attempt to correct for this smearing assuming either a constant or linearly changing radial pulsar acceleration. Such implementations work for integration times, $T \lesssim 0.10P_b$ and $T \lesssim 0.15P_b$, respectively \citep{Ransom2002, Andersen2018}. While acceleration and jerk searches favor shorter $T$ to counter orbital motion, the sensitivity to a single harmonic in the power spectrum also drops with decreasing $T$. Integration times in Fourier-domain pulsar searches must therefore strike a delicate balance between maximizing the single harmonic sensitivity and mitigating power smearing from orbital motion. \\

Here, we leverage $T \in \{5,~30,~60\}$~minutes in the 4--8~GHz Breakthrough Listen (BL) GC survey \citep{Gajjar2021} to conduct sensitive searches for  pulsars orbiting Sgr~A* or stellar-mass BHs. Section~\ref{sec:obs} describes our observations and data preprocessing. In Section~\ref{sec:psearch}, we present our pulsar search methodology and results. We estimate our survey sensitivity to periodic signals in Section~\ref{sec:sensitivity}. Ultimately, we summarize our key findings and discuss their physical significance in Section~\ref{sec:disc}.

% SECTION 2: OBSERVATIONS
\section{Observations} \label{sec:obs}
%%% FIGURE 1
\begin{figure}[t!]
\includegraphics[width=0.48\textwidth]{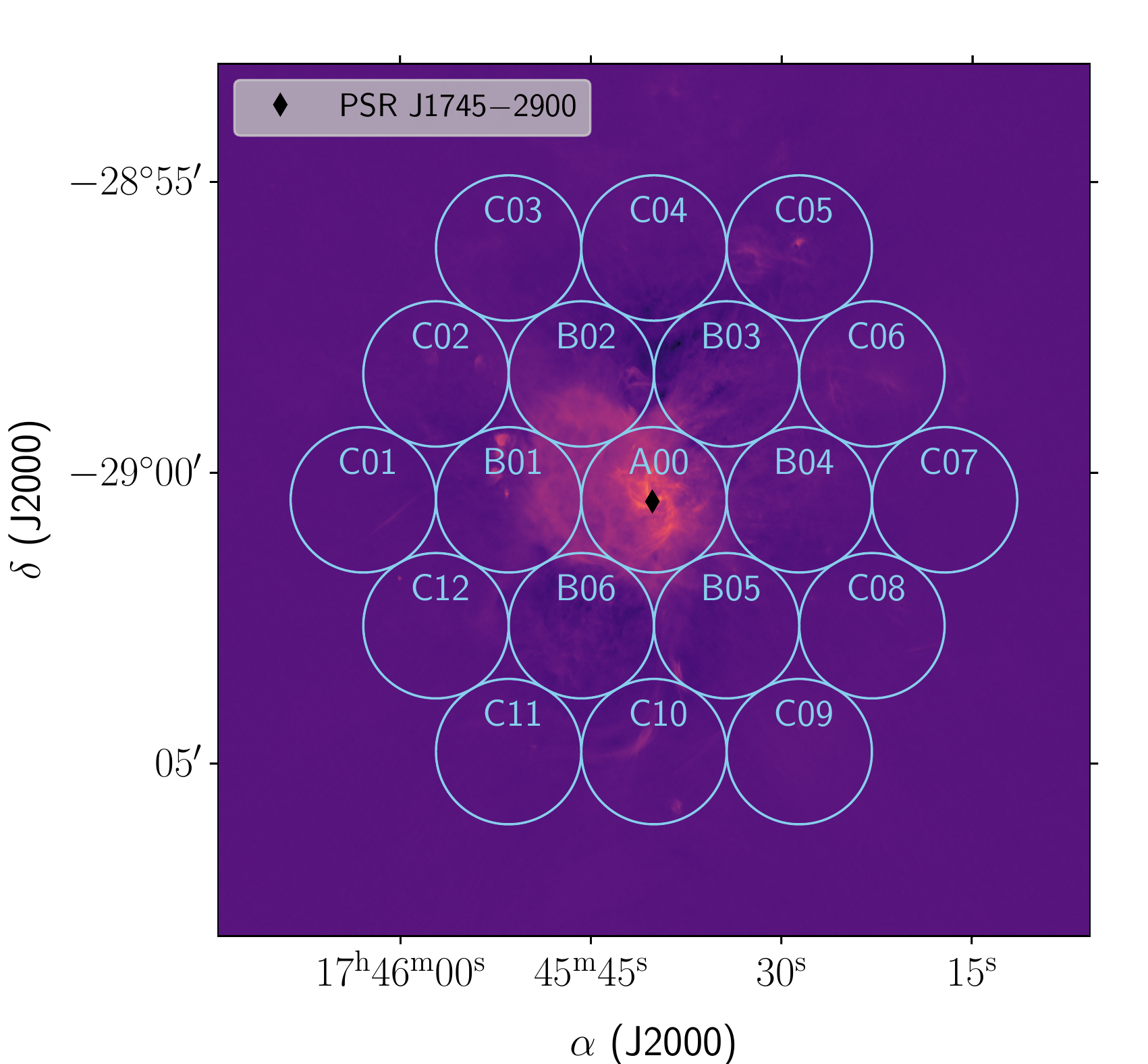}
\caption{4--8~GHz BL GC survey field comprised of 19 distinct pointings (light blue circles) of $\theta_{\rm HPBW} \approx 2\farcm5$ each. The background image is a 5.5~GHz Jansky Very Large Array continuum image \citep{Zhao2016} of the Sgr~A* complex. The GC magnetar PSR~J1745$-$2900 (black diamond) is contained in our central pointing A00. All known pulsars lie $\gtrsim 10\arcmin$ away from Sgr~A*, i.e., outside of our survey footprint. \label{fig1}}
\end{figure}
%%%%%%%%%%%%%%%%%%%%%%%%%%%%%%%%%%
%%%% TABLE 1
\begin{deluxetable*}{CcCcccC}
\tablecaption{Log of 4--8~GHz GBT observations analyzed in our study. \label{tab1}}
\tablewidth{0pt}
\tablehead{
\colhead{Epoch} & \colhead{Start date} & \colhead{Start MJD} & \colhead{Test pulsars} & \colhead{Pointings} & \colhead{$N_{\rm scans}$\tablenotemark{a}} & \colhead{Scan duration} \\
\colhead{(number)} & \colhead{(UTC)} & \colhead{(UTC)} & \colhead{} & \colhead{} & \colhead{} & \colhead{(minutes)}
}
\startdata
1 & 2019 Aug 7 & 58702.217 & B0355$+$54, J1744$–$1134 & (C01,~C07) &  (3,~2) & 5 \\
\hline
2 & 2019 Aug 9 & 58704.993 & B1133$+$16, J1744$–$1134 & C07 & 1 & 5 \\
& & & & (B01,~B04) & (3,~3) & 5 \\
& & & & (B02,~B05) & (3,~3) & 5 \\
& & & & (B03,~B06) & (3,~3) & 5 \\
& & & & (C02,~C04) & (3,~3) & 5 \\
& & & & (C03,~C05) & (3,~3) & 5 \\
& & & & (C06,~C08) & (3,~3) & 5 \\
& & & & (C09,~C11) & (3,~3) & 5 \\
& & & & (C10,~C12) & (2,~3) & 5 \\
& & & & A00 & 1 & 60 \phn \\
\hline
3 & 2019 Sep 8 & 58734.958 & \nodata & A00 & 2 & 30 \phn \\
\hline 
4\tablenotemark{b} &  2019 Sep 11 & 58737.962 & B2021$+$51 & A00 & 8 & 30 \phn \\
\hline
\enddata
\tablenotetext{a}{No. of scans per pointing. For a pointing pair (X,~Y), ($N_{X}$,~$N_{Y}$) denotes the no. of scans of X and Y respectively.}
\tablenotetext{b}{Epoch~4 included position-switched observations of the flux density calibrator 3C~286.}
\tablecomments{Data available for download at the following links. \\ \url{http://blpd9.ssl.berkeley.edu/GCrawspec} \\
\url{http://blpd12.ssl.berkeley.edu/GCrawspec}}
\vspace{-8mm}
\end{deluxetable*}
\vspace{-10mm}
%%%%%%%%%%%%%%%%%%%%%%%%%%%%%%%%%%
The BL GC survey is an extensive 0.7--93~GHz search of the GC and neighboring Galactic bulge fields for radio technosignatures, pulsars, bursts, spectral lines, and masers (see \citealt{Gajjar2021} for the full survey description, data products, and early technosignature and burst science results). The 4--93~GHz component of the survey utilizes the Robert C.~Byrd Green Bank Telescope (GBT), whereas the 0.7--4~GHz portion uses the Parkes radio telescope. Figure~\ref{fig1} shows the 4--8~GHz survey field, wherein a $6\farcm25$ radius of the GC is covered by 19 distinct GBT pointings arranged in three concentric hexagonal rings. From inner to outer, these rings are labeled A, B, and C, with 1, 6, and 12 pointings per ring respectively. All pointings used the single-beam C-band receiver, yielding a half-power beam width, $\theta_{\rm HPBW} \approx 2\farcm5$ at 6~GHz. As illustrated in Figure~\ref{fig1}, our central pointing A00 contains the GC magnetar. We refer readers to \citet{Suresh2021} for a 4--8~GHz study of the GC magnetar using our A00 data. \\

Table~\ref{tab1} presents an overview of our 4--8~GHz observations distributed across four epochs during 2019 August--September. Our observing program consists of eleven deep integrations ($\geq 30$~minutes) on A00, two 5-minutes scans on C10, and three 5-minutes cadences on each of the remaining pointings. In addition, we observed test pulsars at three epochs, and confirmed their respesctive detections to verify our system integrity. To identify and reject radio frequency interference (RFI) via position switching, we conducted alternating observations of pairs of pointings in rings B and C. Pointing pairs were chosen such that the beam centers of grouped pointings were separated by at least $2\theta_{\rm HPBW}$ on the sky. \\

To accommodate various science cases, baseband voltages gathered during our observations were channelized to different spectral and temporal resolutions using the Breakthrough Listen Digital Backend \citep{MacMahon2018,Lebofsky2019}. Here, for our pulsar searches, we worked with total intensity filterbank data (no coherent dedispersion performed) having $\approx 43.69~\mu$s time sampling and $\approx 91.67$~kHz channel bandwidth. These data contain 53,248 channels spanning 3.56--8.44~GHz, which covers the 3.9--8.0~GHz instantaneous response of the C-band receiver.

\subsection{Data preprocessing} \label{sec:preprocessing}
We followed the methodology of \citet{Suresh2021} to excise RFI from our data using the {\tt rfifind} module of the pulsar software package {\tt PRESTO} \citep{PRESTO}. Adopting an integration time of 1~s for our {\tt rfifind} runs, we detected bright, persistent interference between 4.24--4.39, 4.90--4.95, and 6.90--7.10~GHz. Incorporating our {\tt rfifind} mask and clipping bandpass edges, the usable radio frequency band in our data extends between 4.4--8.0~GHz. \\

After RFI masking, we dedispersed our dynamic spectra (radio frequency-time data) at 1836 trial DMs between 0~pc~cm$^{-3}$ and 5505~pc~cm$^{-3}$ (both limits included) with a grid spacing of 3~pc~cm$^{-3}$. These dedispersed data were summed over the entire usable band, and then block-averaged by a factor of 8 to output dedispersed time series with a sample interval, $t_{\rm samp} \approx 349.53~\mu$s.

\subsection{Red noise removal} \label{sec:rednoise}
Slow pulsar discovery ($P_0 \geq 1$~s) in long time series ($T \geq 5$~minutes) often suffers from the presence of low-frequency noise in Fourier-domain spectra. To alleviate the adverse impact of red noise on our pulsar searches, we detrended our dedispersed time series using a running median window of width $W_{\rm med}$. \\

For deciding an optimal value of $W_{\rm med}$, we visually inspected the effect of  different trial $W_{\rm med}$ on power spectra of barycentric GC magnetar time series (${\rm DM} = 1776$~pc~cm$^{-3}$, $f_{0}^{\rm mag} = 1/P_{0}^{\rm mag}\approx 0.2653$~Hz, \citealt{Suresh2021}). Starting at $W_{\rm med} = 4$~s, we successively lowered $W_{\rm med}$ by factors of 2 until further reduction in $W_{\rm med}$ brought no concomitant increase in the number of harmonics of $f_{0}^{\rm mag}$ seen in the power spectrum. In doing so, we settled at $W_{\rm med} = 0.25$~s for removing slow baseline fluctuations in our dedispersed time series. \\
%%% FIGURE 2
\begin{figure}[t!]
\includegraphics[width=0.48\textwidth]{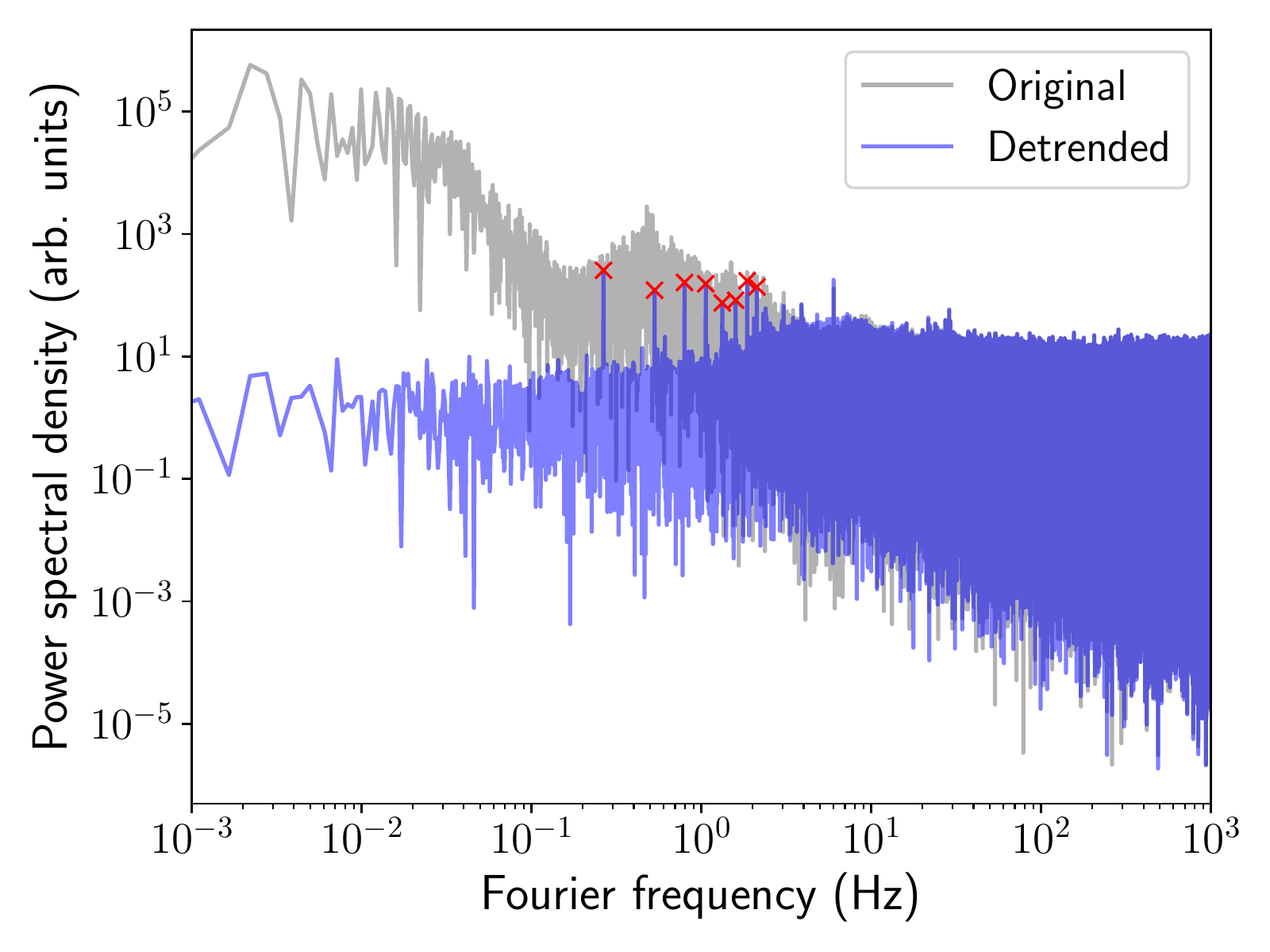}
\caption{Power spectra of barycentric ${\rm DM} = 1776$~pc~cm$^{-3}$ time series from a 30-minutes A00 scan before (gray) and after (blue) detrending with a running median filter of width 0.25~s. The red crosses label the fundamental rotation frequency of the GC magnetar and its harmonics. The red noise in the gray power spectrum below 4~Hz arises from gradual baseline fluctuations in our original time series. \label{fig2}}
\end{figure}
%%%%%%%%%%%%%%%%%%%%%%%%%%%%%%%%%%

Figure~\ref{fig2} shows the result of time series detrending on the power spectrum of a barycentric ${\rm DM} = 1776$~pc~cm$^{-3}$ time series from a 30-minutes A00 scan. The power spectrum of the detrended time series evidently reveals significant peaks at $f_{0}^{\rm mag}$ and its first seven harmonics. Without detrending, these peaks remain buried within red noise in the power spectrum of the original time series. \\

Finally, to limit the false positive count in our pulsar searches, we masked periodic RFI in the power spectra of our detrended time series. Looking at power spectra of topocentric ${\rm DM} = 0$~pc~cm$^{-3}$ time series, we identified and flagged significant spikes at frequencies of 1.2, 6, 60~Hz (US electric power line frequency), and their first five harmonics. The resulting cleaned power spectra constituted the basis of our Fourier-domain periodicity searches.

% SECTION 3: PERIODICITY SEARCHES
\section{Periodicity Searches} \label{sec:psearch}
%%% FIGURE 3
\begin{figure}[t!]
\includegraphics[width=0.48\textwidth]{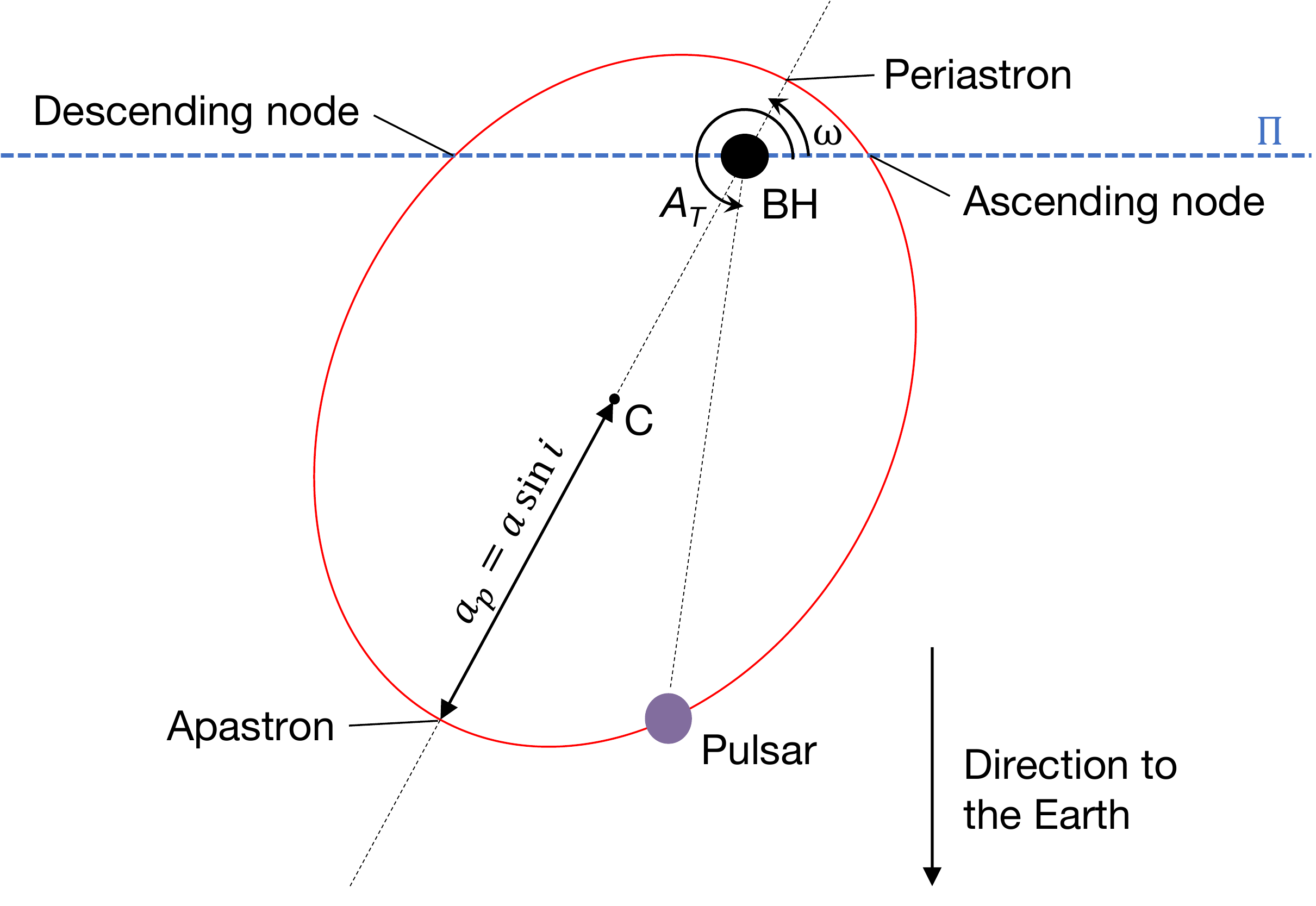}
\caption{A pulsar orbiting a black hole. The pulsar orbit (red ellipse) has been projected onto a plane containing the direction to the Earth and the line $\Pi$ connecting the two nodes. The symbols $a$, $i$, $\omega$, and $A_T$ represent, respectively, the orbital semi-major axis, the orbital inclination, the argument of periapsis, and the pulsar true anomaly. The quantity $a_p = a \sin{i}$ is the projected semi-major axis of the pulsar orbit centered at C. \label{fig3}}
\end{figure}
%%%%%%%%%%%%%%%%%%%%%%%%%%%%%%%%%%
%%% FIGURE 4
\begin{figure*}[ht!]
\includegraphics[width=\textwidth]{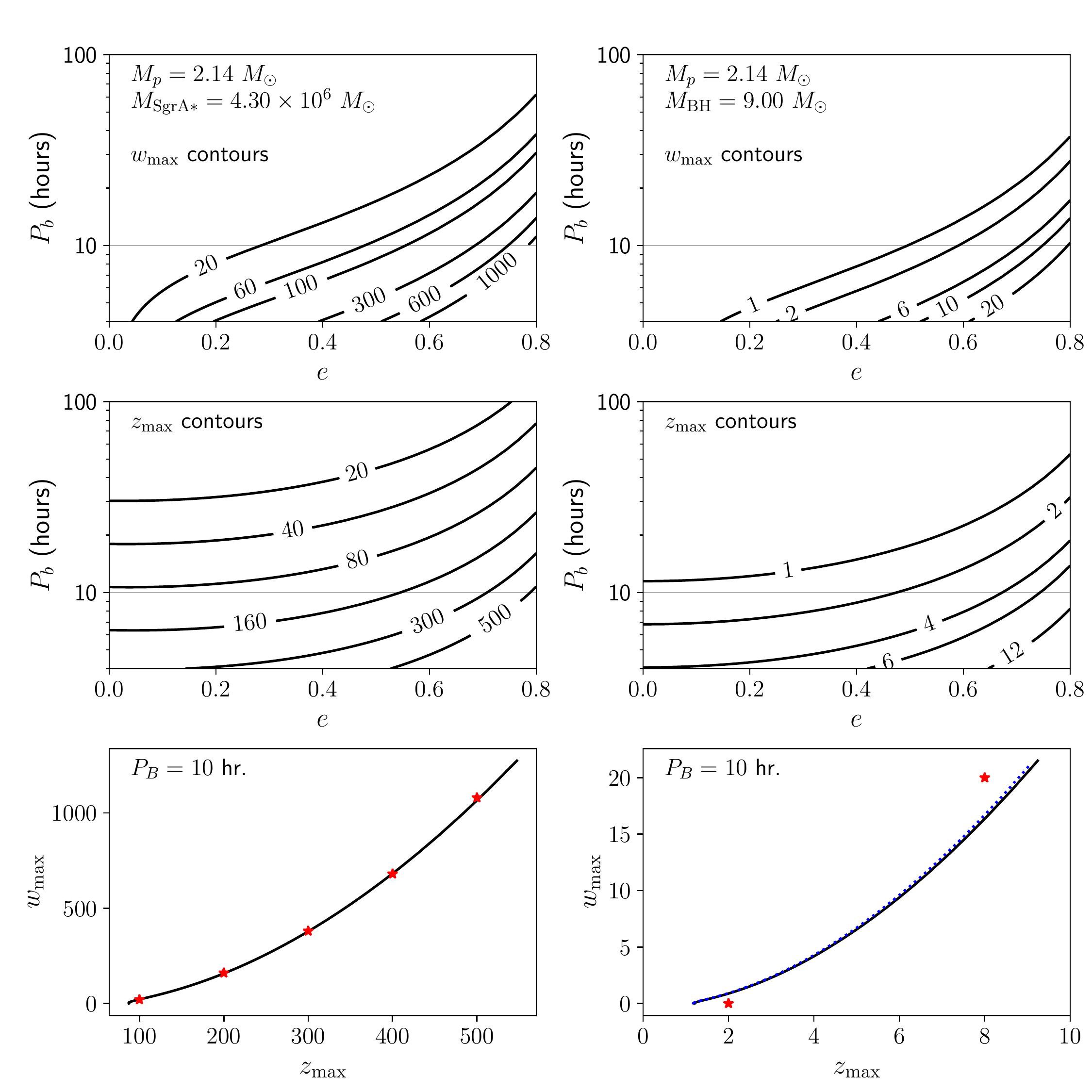}
\caption{($z_{\rm max}$,~$w_{\rm max}$) curves for $N_h=1$ binary pulsar searches. Top row: Contours of constant $w_{\rm max}$ for a $2.14~M_{\odot}$ pulsar orbiting Sgr~A* (left column) and a $9~M_{\odot}$ BH (right column). Middle row: $z_{\rm max}$ contours for the same systems. Bottom row: $w_{\rm max}$ vs. $z_{\rm max}$ (black solid curves) for a binary orbital period of 10~hours, assuming a pulsar mass of $2.14~M_{\odot}$. While the black solid curve in the bottom left panel is insensitive to $M_p$, the blue dotted curve in the bottom right panel shows the corresponding trend for a $1.4~M_{\odot}$ pulsar orbiting a $9~M_{\odot}$ BH. The red stars mark ($z_{\rm max}$,~$w_{\rm max}$) tuples chosen for our Fourier-domain jerk searches on $T=30$~min. integrations. We note that the {\tt accelsearch} module of {\tt PRESTO} quantizes $z_{\rm max}$ and $w_{\rm max}$ in multiples of 2 and 20 respectively \citep{Andersen2018}. All panels assume $P_0 = 1$~s, and edge-on pulsar orbits with $\omega = 0\degr$ and $A_T = -90\degr$. 
\label{fig4}}
\end{figure*}
%%%%%%%%%%%%%%%%%%%%%%%%%%%%%%%%%
Consider a regular pulse train (insignificant pulse jitter) from an isolated pulsar of barycentric rotational frequency, $f_0 = 1/P_0$. For an effective pulse duty cycle $\delta_{\rm eff}$, the energy in the pulse train gets distributed over $N_h \simeq 1/2\delta_{\rm eff}$ independent harmonics in the power spectrum. Consequently, standard periodicity searches perform harmonic summing in power spectra of dedispersed time series to increase the significance of a pulsar detection \citep{Ransom2002}. In a blind search, the pulsar DM, $f_0$, and $\delta_{\rm eff}$ are unknown apriori. Searches for isolated pulsars, hence require sampling of the three-dimensional parameter space DM--$f$--$N_h$. \\

Binary orbital motion complicates pulsar detection by introducing a time-dependent Doppler drift that smears harmonics in the power spectrum. Conventional search algorithms attempt to retroactively correct for this smearing by presuming a constant or linearly evolving line-of-sight pulsar acceleration \citep{Ransom2002,Andersen2018}. \\

Figure~\ref{fig3} shows a binary pulsar orbit and its Keplerian orbital elements, including its projected semi-major axis $a_p = a\sin{i}$ (orbital semi-major axis $a$ and inclination $i$), argument of periapsis $\omega$, and true anomaly $A_T$. In the Newtonian regime, the line-of-sight pulsar acceleration ($a_l$) and jerk ($j_l$) are respectively given by \citep{Bagchi2013,Liu2021}
%%% EQUATION 1
\begin{align} \label{eqn1}
a_l = &- \left( \frac{2\pi}{P_b} \right)^2 \frac{a_p}{(1-e^2)^2} \left(1+e\cos{A_T}\right)^2 \times \nonumber \\
& \sin{(A_T + \omega)},
\end{align}
%%%%%%%%%%%%%%%%%%%%%%%%%%%%%%%%%
%%% EQUATION 2
\begin{align} \label{eqn2}
j_l = &- \left( \frac{2\pi}{P_b} \right)^3 \frac{a_p}{(1-e^2)^{7/2}} \left(1+e\cos{A_T}\right)^3 [e\cos{\omega} \nonumber \\
& + \cos{(A_T + \omega)} - 3e\sin{(A_T + \omega)}\sin{A_T}].
\end{align}
%%%%%%%%%%%%%%%%%%%%%%%%%%%%%%%%%

In the above equations, $e$ denotes the eccentricity of the pulsar orbit. Casting $a_l$ and $j_l$ in dimensionless units, and introducing the harmonic number $h$, we have
%%% EQUATION 3
\begin{align} \label{eqn3}
z &= h \left(\frac{a_lfT^2}{c} \right),
\end{align}
%%%%%%%%%%%%%%%%%%%%%%%%%%%%%%%%%
%%% EQUATION 4
\begin{align} \label{eqn4}
w &= h \left(\frac{j_lfT^3}{c} \right).
\end{align}
%%%%%%%%%%%%%%%%%%%%%%%%%%%%%%%%%

%%%% TABLE 2
\begin{deluxetable*}{lcCCl}
\tablecaption{Pulsar search parameters \label{tab2}}
\tablewidth{0pt}
\tablehead{
\colhead{Science targets} & \colhead{Pointings} & \colhead{$T$\tablenotemark{$\dagger$}} & \colhead{$P_b$\tablenotemark{$\ddagger$}} & \colhead{($z_{\rm max}$, $w_{\rm max}$)} \\
\colhead{} & \colhead{} & \colhead{(minutes)} & \colhead{(hours)} & \colhead{}
}
\startdata
Isolated pulsars & All & 5,~30,~60 & \nodata & (0,~0) \\
\hline
Pulsars orbiting stellar-mass BHs & B-ring, C-ring & 5 & 1 & (2,~0), (6,~20) \\
& A00 & 30 & 10 & (2,~0), (8,~20) \\
\hline
Pulsars around Sgr~A* & A00 & 30 & 10 & (100,~20), (200,~160), \\ 
& & & & (300,~380), (400,~680), \\
& & & & (500,~1080) \\
\hline
\enddata
\tablenotetext{\dagger}{Integration time}
\tablenotetext{\ddagger}{Target binary orbital period assumed for specified ($z_{\rm max}$, $w_{\rm max}$) tuples}
\tablecomments{
\begin{enumerate}[leftmargin=*]
\item{1836 trial DMs explored between 0--5505~pc~cm$^{-3}$ (both limits included) with a grid spacing of 3~pc~cm$^{-3}$.}
\item{For a sample interval of $\approx 349.53~\mu$s in the dedispersed time series, the range of $f$ searched is 0--1430~Hz with a resolution of $1/T$.}
\item{Trial $N_h$ values considered $=$ \{1, 2, 4, 8\}.}
\end{enumerate}
}
\vspace{-10mm}
\end{deluxetable*}
\vspace{-5mm}
%%%%%%%%%%%%%%%%%%%%%%%%%%%%%%%%%%

Here, $h=1$ labels the fundamental frequency $f$, and $c$ is the vacuum speed of light. Following \citet{Andersen2018}, we define the dimensionless Fourier frequency, $r=fT$. The quantities $z$ and $w$ thus represent the number of bins of signal drift in $r$ and $\dot{r}$ over time $T$. In Equations~\ref{eqn3} and \ref{eqn4}, we note that both $z$ and $w$ scale linearly with $h$. Therefore, orbital motion lowers the detection significance of all harmonics in the power spectrum, with greater deleterious effects for higher harmonics. The most minimal regime of pulsar detection then occurs when only the fundamental survives with adequate significance in the power spectrum, while all higher harmonics have been smeared into the continuum by orbital motion.

\subsection{Acceleration and jerk searches}\label{sec:acc_jerk}
The {\tt accelsearch} routine of {\tt PRESTO} \citep{PRESTO} executes a matched filtering scheme that accounts for Fourier-domain power smearing of the highest harmonic up to a Fourier acceleration $z_{\rm max}$ and a Fourier jerk $w_{\rm max}$ \citep{Ransom2002,Andersen2018}. Searches for binary pulsars, hence involve sampling over five parameters, i.e., the DM, $f$, $N_h$, $z_{\rm max}$, and $w_{\rm max}$. \\

We targeted the discovery of pulsars in compact orbits around Sgr~A* or stellar-mass BHs. We focused our searches on CPs as we expected most MSPs to fall below our detection threshold (see Section~\ref{sec:sensitivity}). To identify suitable trial parameter values for our binary pulsar searches, we considered a $M_p = 2.14~M_{\odot}$ pulsar (highest neutron star mass measured to date, \citealt{Cromartie2020}) located at $A_T=-90\degr$ in an edge-on orbit ($i=90\degr$) with $\omega=0\degr$. We further envisaged $N_h=1$ searches for pulsars in tight binaries with large $a_l$ and $j_l$. \\

Assuming $T=30$~minutes and $P_0 = 1$~s, Figure~\ref{fig4} shows contours of constant $w_{\rm max}$ (top row) and $z_{\rm max}$ (middle row) for pulsar orbits around Sgr~A* (left column) and a $M_{\rm BH} = 9~M_{\odot}$ BH (median BH mass for solar metallicity, \citealt{Woosley2020}). The condition $T \lesssim 0.15 P_b$ for jerk searches restricts us to $P_b \gtrsim 3.3$~hours for $T = 30$~minutes. Setting $P_b = 10$~hours as our target, we sought to cover $e \in \left[ 0.0,~0.8 \right]$ by uniformly sampling the $w_{\rm max}$ vs. $z_{\rm max}$ curves shown in the bottom row of Figure~\ref{fig4}. For our selected $M_{\rm BH}$ value, $a_p$ varies inappreciably for known ranges of $M_p$. Therefore, our chosen ($z_{\rm max},~w_{\rm max}$) tuples encompass all known pulsar masses, including the typical mass of $1.4~M_{\odot}$ \citep{Zhang2011}. \\

%%%% TABLE 3
\begin{deluxetable}{cCCC}
\tablecaption{Periodicity detection statistics \label{tab3}}
\tablewidth{0pt}
\tablehead{
\colhead{Pointing} & \colhead{$N_{\rm detected}$\tablenotemark{a}} & \colhead{$N_{\rm rejected}$\tablenotemark{b}} & \colhead{$N_{\rm cands}$\tablenotemark{c}}
}
\startdata
A00 & 7039 \phn &\nodata & 7039 \phn \\
\hline
B01 & 454 & 27 & 427 \\
B02 & 487 & 22 & 465 \\
B03 & 495 & 26 & 469 \\
B04 & 460 & 34 & 426 \\
B05 & 414 & 25 & 389 \\
B06 & 485 & 25 & 460 \\
\hline
C01 & 481 & 13 & 468 \\
C02 & 456 & 27 & 429 \\
C03 & 489 & 36 & 453 \\
C04 & 437 & 25 & 412 \\
C05 & 554 & 28 & 526 \\
C06 & 596 & 48 & 548 \\
C07 & 458 & 12 & 446 \\
C08 & 423 & 33 & 390 \\
C09 & 555 & 46 & 509 \\
C10 & 370 & 12 & 358 \\
C11 & 413 & 38 & 375 \\
C12 & 982 & 11 & 971 \\
\hline
Total & 16048 \phn \phn & 488 \phn & 15560 \phn \phn \\
\enddata
\tablenotetext{a}{No. of distinct candidates detected at or above $6\sigma_{\rm ps}$ equivalent Gaussian significance in $\chi^2$-distributed power spectra.}
\tablenotetext{b}{No. of candidates rejected via position switching.}
\tablenotetext{c}{No. of candidates that pass comparisons between periodicity detections in paired pointings.}
\vspace{-10mm}
\end{deluxetable}
\vspace{-5mm}
%%%%%%%%%%%%%%%%%%%%%%%%%%%%%%%%%%

Table~\ref{tab2} outlines our pulsar search strategy, wherein we also incorporated searches for isolated pulsars and binary pulsar orbits in the $T=5$~minutes integrations of our B- and C-ring pointings. Again, we identified viable $z_{\rm max}$ and $w_{\rm max}$ values considering $N_h=1$ searches for binary pulsar orbits. Adopting the same ($z_{\rm max}$,~$w_{\rm max}$) tuples for our $N_h = \{2,~4,~8\}$ searches, we enhance our sensitivity to slowly orbiting pulsars through harmonic summing. However, since our searches perform matched filtering \citep{Andersen2018} in the Fourier-domain, greatest sensitivity will be had for systems with $a_l = (z_{\rm max} c/N_hf T^2)$ and $j_l = (w_{\rm max} c/ N_h fT^3)$. \\

Our pulsar search procedure on data can be described in three independent stages. Firstly, we ran non-accelerated searches for isolated pulsars on a per-scan basis. Secondly, at the trial parameter values listed in Table~\ref{tab2}, we conducted acceleration and jerk searches on data from our 5-minutes and 30-minutes scans. Finally, we split our 60-minutes A00 scan from epoch~2 into two contiguous halves, and executed jerk searches separately on each half. \\

Imposing a $6\sigma_{\rm ps}$ detection probability threshold in harmonic-summed power spectra, we used the {\tt accel$\_$sift.py} script of {\tt PRESTO} \citep{PRESTO} to group hits across adjacent trial DMs and frequencies $f$ into distinct candidates. Here, $\sigma_{\rm ps}$ measures the equivalent Gaussian significance of a frequency $f$ in the $\chi^2$-distributed power spectrum (assuming Gaussian white noise background in detrended time series). 

\subsection{Results} \label{sec:results}
Our pulsar searches yielded a total of 16,048 periodicity candidates across 64 scans. Table~\ref{tab3} presents our candidate detection statistics organized by pointings. Since we expect any astrophysical signal of our interest to be localized on the sky, we pruned our detection list by rejecting candidates common to paired pointings within a frequency tolerance $\Delta f$ and a DM tolerance $\Delta$DM. For an integration time $T$, we chose $\Delta f = 1/2T$, i.e., the maximum error on $f$ for $N_h = 1$. Given an effective pulse width $W_{\rm eff} = \delta_{\rm eff}P$, the DM uncertainty associated with pulse detection across a usable bandwidth $B$ is \citep{Cordes2003}
%%% EQUATION 5
\begin{align}\label{eqn5}
\Delta {\rm DM} & \simeq 506~{\rm pc~cm}^{-3} \left( \frac{W_{\rm eff,ms} \nu_{c, {\rm GHz}}^3}{B_{\rm MHz}} \right).
\end{align}
%%%%%%%%%%%%%%%%%%%%%%%%%%%%%%%%%
Here, $B \approx 3.35$~GHz, and $\nu_c \approx 6.1$~GHz is the center frequency of our observations. Conservatively setting $W_{\rm eff} = 3t_{\rm samp} \approx 1$~ms, we obtain $\Delta {\rm DM} \approx 33$~pc~cm$^{-3}$. As evident from Table~\ref{tab3}, our position-switched observations permit us to eliminate about $5\%$ of candidates in rings B and C, thereby leaving us with 15,560 periodicity detections across 64 scans. \\

%%% FIGURE 5
\begin{figure*}[t!]
\includegraphics[width=\textwidth]{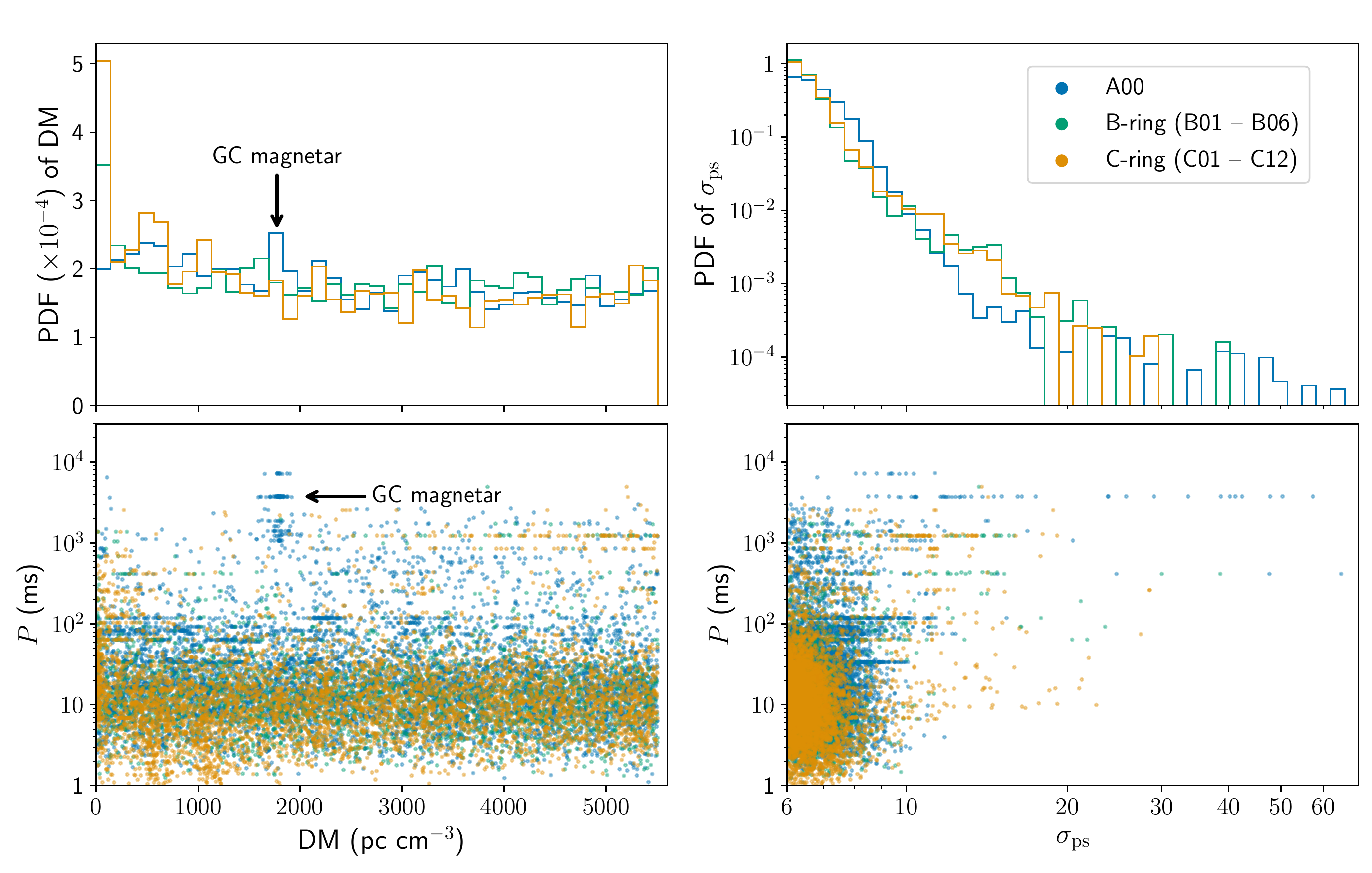}
\caption{Top left panel: Probability distribution function (PDF) of candidate DMs binned uniformly to 141~pc~cm$^{-3}$ resolution. Top right panel: PDF of equivalent Gaussian significance ($\sigma_{\rm ps}$) of candidates in power spectra. Plotted bins are linear in $\ln \sigma_{\rm ps}$ with width $\approx 0.06$. Bottom left panel: Scatter of candidates in the period--DM plane. Bottom right panel: Period--$\sigma_{\rm ps}$ scatter of candidates. In all panels, the blue, green, and orange colors represent A00, B-ring, and C-ring respectively. \label{fig5}}
\end{figure*}
%%%%%%%%%%%%%%%%%%%%%%%%%%%%%%%%%
%%% FIGURE 6
\begin{figure}[t!]
\includegraphics[width=0.48\textwidth]{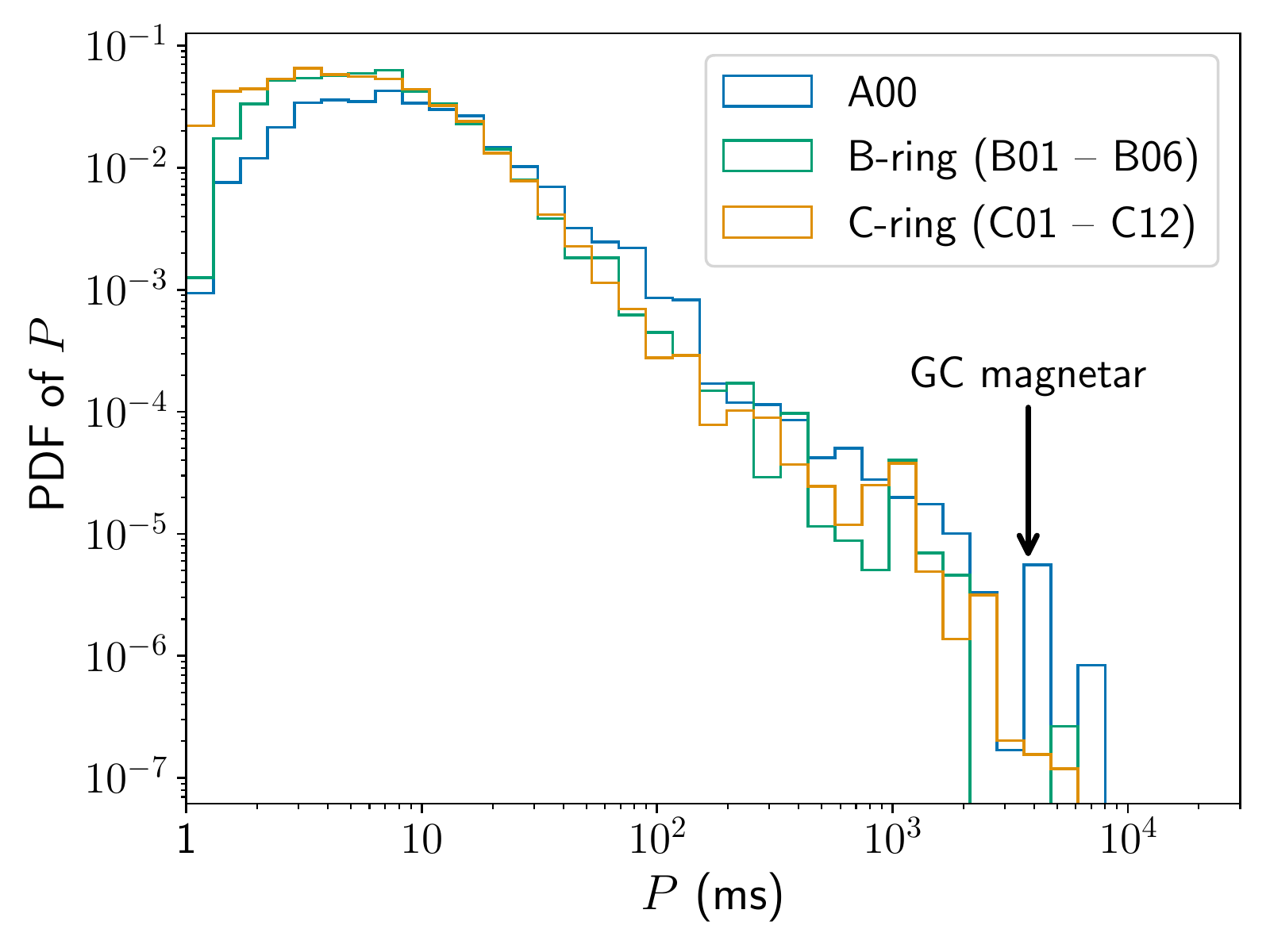}
\caption{PDF of candidate periods ($P$) organized by rings. Histogram bins are evenly spaced in $\ln P$ with width $\approx 0.26$~s. \label{fig6}}
\end{figure}
%%%%%%%%%%%%%%%%%%%%%%%%%%%%%%%%%
In Figures~\ref{fig5} and \ref{fig6}, we statistically analyze the remaining 15,560 candidates in terms of their DM, period ($P = 1/f$), and $\sigma_{\rm ps}$ distributions. The GC magnetar markedly stands out as a high significance cluster of A00 candidates centered at ${\rm DM} = 1776$~pc~cm$^{-3}$ and $P = 3.7686$~s. Furthermore, nearly $90\%$ of candidates fall within a horizontal expanse between $P = 2$~ms and $P=100$~ms in the $P$--DM plane. We attribute this scatter of candidates partly to a statistical sampling effect (greater number of power spectrum samples across $P \in [2,~100]$~ms than that spanning $P \in [10^2,~10^4]$~ms), and partly to noise and intermittent band-limited periodic RFI in dynamic spectra. Appendix~\ref{appendix:avg_profiles} shows average profiles of two sample candidates, whose phase-resolved dynamic spectra reveal bright RFI between 4.4--4.9, 5.1--5.2, 6.8--7.0, and 7.5--7.8~GHz. These narrowband structures asynchronously wax and wane during our observations, leading to a horde of non-astrophysical candidates detected at numerous DMs and periods.\\

%%% TABLE 4
\begin{deluxetable*}{lCCCCCCCC}
\tablecaption{Observing parameters and sensitivity thresholds for GC pulsar surveys conducted at 4--9~GHz \label{tab4}}
\tablewidth{\textwidth}
\tablehead{
\colhead{Survey} & \colhead{$\nu_c$} & \colhead{$B$} & \colhead{$T$} & 
\colhead{$t_{\rm samp}$} &
\colhead{$S_{\rm sys} (\nu_c)$\tablenotemark{a}} &
\colhead{$L_{\rm sh} (\nu_c)$\tablenotemark{b}} &
\colhead{$L_{\rm min} (\nu_c)$ \tablenotemark{c}} &
\colhead{$L_{\rm min}$ (6.1~GHz) \tablenotemark{d}}
\\
\colhead{} & \colhead{(GHz)} & \colhead{(GHz)} & \colhead{(min.)} & \colhead{(ms)} & \colhead{(Jy)} & \colhead{(mJy kpc$^2$)} & \colhead{(mJy kpc$^2$)} & \colhead{(mJy kpc$^2$)}
}
\startdata
\citet{Johnston2006} & 8.4 & 0.864 & 70 & 1.0 & 148 & 22.1 & 3.5 & 5.5 \\
\citet{Deneva2009} & 4.8 & 0.8 & 60 & 0.65 & \phn 57.5\tablenotemark{e} & 9.6 & 1.5 & 1.1 \\
\citet{Bates2011} & 6.59 & 0.576 & 280 & 0.13 & 216.7 & 19.8 & 3.2 & 3.5 \\
\citet{Eatough2021} & 4.85 & 0.5 & 72 & 0.26 & 129 & 24.9 & 4.0 & 2.9 \\
& 8.35 & 0.5 & 144 & 0.13 & 93.3 & 12.7 & 2.0 & 3.2 \\
\hline
This work\tablenotemark{f}: & & & & & \\
A00 & 6.1 & 3.35 & 30 & 0.35 & 46.3 & 5.3 & 0.9 & 0.9\\
Off-Sgr~A* pointings\tablenotemark{g} & 6.1 & 3.35 & 5 & 0.35 & 10.7 & 3.0 & 0.5 & 0.5
\enddata
\tablenotetext{a}{$S_{\rm sys} = S_{\rm sys}^{\rm off-GC} + S_{\rm sys}^{\rm SgrA*}$.}
\tablenotetext{b}{Single harmonic pseudo-luminosity threshold, $L_{\rm sh} = S_{\rm sh}d_{\rm GC}^2$, where $d_{\rm GC} = 8.18$~kpc \citep{GRAVITY2019}.}
\tablenotetext{c}{Minimum detectable pseudo-luminosity, $L_{\rm min} (\nu_c) = S_{\rm min} (\nu_c) d_{\rm GC}^2$ for $\delta_{\rm int} = 2.5\%$ and $P_0 = 1$~s.}
\tablenotetext{d}{$L_{\rm min} \propto \nu^{\alpha}$ invoked, where $\alpha = -1.4$ is the mean pulsar spectral index \citep{Bates2013}.}
\tablenotetext{e}{$T_{\rm GC}(\nu) \approx 568~{\rm K} (\nu/1~{\rm GHz})^{-1.13}$ \citep{Rajwade2017} assumed to compute $S_{\rm sys}^{\rm SgrA*}$.}
\tablenotetext{f}{We estimated $S_{\rm sys}^{\rm off-GC}$ and $S_{\rm sys}^{\rm SgrA*}$ using Figure~1 and Equation~2 of \citet{Suresh2021}.}
\tablenotemark{g}{B-ring and C-ring pointings with negligible $S_{\rm sys}^{\rm SgrA*}$ in Figure~\ref{fig1}.}
\end{deluxetable*}
\vspace{-7mm}
%%%%%%%%%%%%%%%%%%%%%%%%%%%%%%%%%

Other notable features in Figure~\ref{fig5} include streaks of A00 candidates at $P \in [30,~100]$~ms, and an excess of C-ring candidates at $P \lesssim 2$~ms and ${\rm DM} \in [0,~150]\cup[430,700]\cup[1000,~1150]$~pc~cm$^{-3}$. Upon manual visual inspection of their respective folded profiles, we traced the A00 streaks to periodic irregularities occurring near the start and end of individual scans. Further, we associate the surfeit of C-ring candidates with the emergence of bright RFI at 5.8--5.9 and 7.8--7.9~GHz during our C12 scans. \\

In summary, we report the successful detection of the GC magnetar in periodicity searches of our A00 scans. Setting a 6$\sigma_{\rm ps}$ detection threshold for our Fourier-domain searches, we find no statistical evidence of novel periodic astrophysical emissions in the 4--8~GHz BL GC survey data.

% SECTION 4: SURVEY SENSITIVITY
\section{Survey Sensitivity Estimation} \label{sec:sensitivity}
Consider the minimal detection of a pulsar, with only its $h=1$ harmonic visible in the power spectrum. In such a circumstance, the relevant theoretical detection limit is the single harmonic sensitivity at $f_0$ given by
%%% EQUATION 6
\begin{align}\label{eqn6}
S_{\rm sh} = m \frac{S_{\rm sys}}{\sqrt{2BT}}. 
\end{align}
Here, $S_{\rm sys}$ is the system-equivalent flux density (SEFD), and $m = 6$ is the minimum significance required to claim a detection. Often, $S_{\rm sys}$ in GC surveys is written as a sum of two terms.
%%% EQUATION 7
\begin{align}\label{eqn7}
S_{\rm sys}(\nu) = S_{\rm sys}^{\rm off-SgrA*}(\nu) + S_{\rm sys}^{\rm SgrA*}(\nu),    
\end{align}
%%%%%%%%%%%%%%%%%%%%%%%%%%%%%%%%%
where $S_{\rm sys}^{\rm off-SgrA*}$ is the SEFD away from the Sgr~A* complex, and $S_{\rm sys}^{\rm SgrA*}$ captures the SEFD contribution from the Sgr~A* complex. \\

Assuming slow pulsar orbital motion in time $T$, we can lower our sensitivity threshold below $S_{\rm sh}$ by summing over harmonics in the power spectrum. For $\delta_{\rm eff} \ll 1$, the corresponding minimum detectable flux density \citep{Dewey1985} is then
%%% EQUATION 8
\begin{align}\label{eqn8}
S_{\rm min} = m \frac{S_{\rm sys}}{\sqrt{2BT}} \left(\frac{\delta_{\rm eff}}{1-\delta_{\rm eff}} \right)^{1/2}.   
\end{align}
%%%%%%%%%%%%%%%%%%%%%%%%%%%%%%%%%
Appendix~\ref{appendix:Weff} describes various sources of pulse broadening that can widen a pulse from intrinsic width $W_{\rm int}$ to effective width, $W_{\rm eff} = \delta_{\rm eff}P$ in a dedispersed time series. We define $\delta_{\rm int} = W_{\rm int}/P$ as the intrinsic pulse duty cycle. From the ATNF pulsar catalog v1.66 \citep{Manchester2001}\footnote{\url{https://www.atnf.csiro.au/research/pulsar/psrcat}}, about $83\%$ of CPs have duty cycles between 1$\%$ and 10$\%$, with a median duty cycle of $2.5\%$.\\

Assuming negligible instrumental and dispersive pulse broadening, 
%%% EQUATION 9
\begin{align}\label{eqn9}
W_{\rm eff}(\nu) \approx \left(W_{\rm int}^2 + t_{\rm samp}^2 + \tau_{\rm sc}^2(\nu) \right)^{1/2}.  
\end{align}
%%%%%%%%%%%%%%%%%%%%%%%%%%%%%%%%%
Here, $\tau_{\rm sc}(\nu)$ is the scatter-broadening time scale. In the absence of alternate observational evidence, we take $\tau_{\rm sc}$ towards the GC magnetar to be representative of the central ISM throughout the remainder of our study. From \citet{Spitler2014}, we have
%%% EQUATION 10
\begin{align}\label{eqn10}
\tau_{\rm sc}(\nu) \simeq (1.3~{\rm s})~\nu_{c,\rm GHz}^{-3.8},    
\end{align}
%%%%%%%%%%%%%%%%%%%%%%%%%%%%%%%%%
for GC magnetar pulses. Thus, scattering inhibits the detection of GC pulsars with $P_0 \leq \tau_{\rm sc}(6.1~{\rm GHz}) \simeq 1.35$~ms. \\

Table~\ref{tab4} lists observational parameters and theoretical sensitivity limits of GC pulsar surveys conducted at $\nu_c$ between 4--9~GHz. We adopted a GC distance, $d_{\rm GC} = 8.18$~kpc \citep{GRAVITY2019} to convert $S_{\rm sh}$ and $S_{\rm min}$ to psuedo-luminosities, $L_{\rm sh}$ and $L_{\rm min}$, respectively. To facilitate sensitivity comparison across surveys at different $\nu_c$, we scaled their respective $L_{\rm min}$ to 6.1~GHz invoking $L_{\rm min} \propto \nu^{\alpha} $, where $\alpha = -1.4$ is the mean pulsar spectral index \citep{Bates2013}. This scaling preserves the fraction of known pulsars with spectral pseudo-luminosity, $L_{\nu} \geq L_{\rm min}(\nu)$ for a given survey. \\  

%%% FIGURE 7
\begin{figure}[t!]
\includegraphics[width=0.48\textwidth]{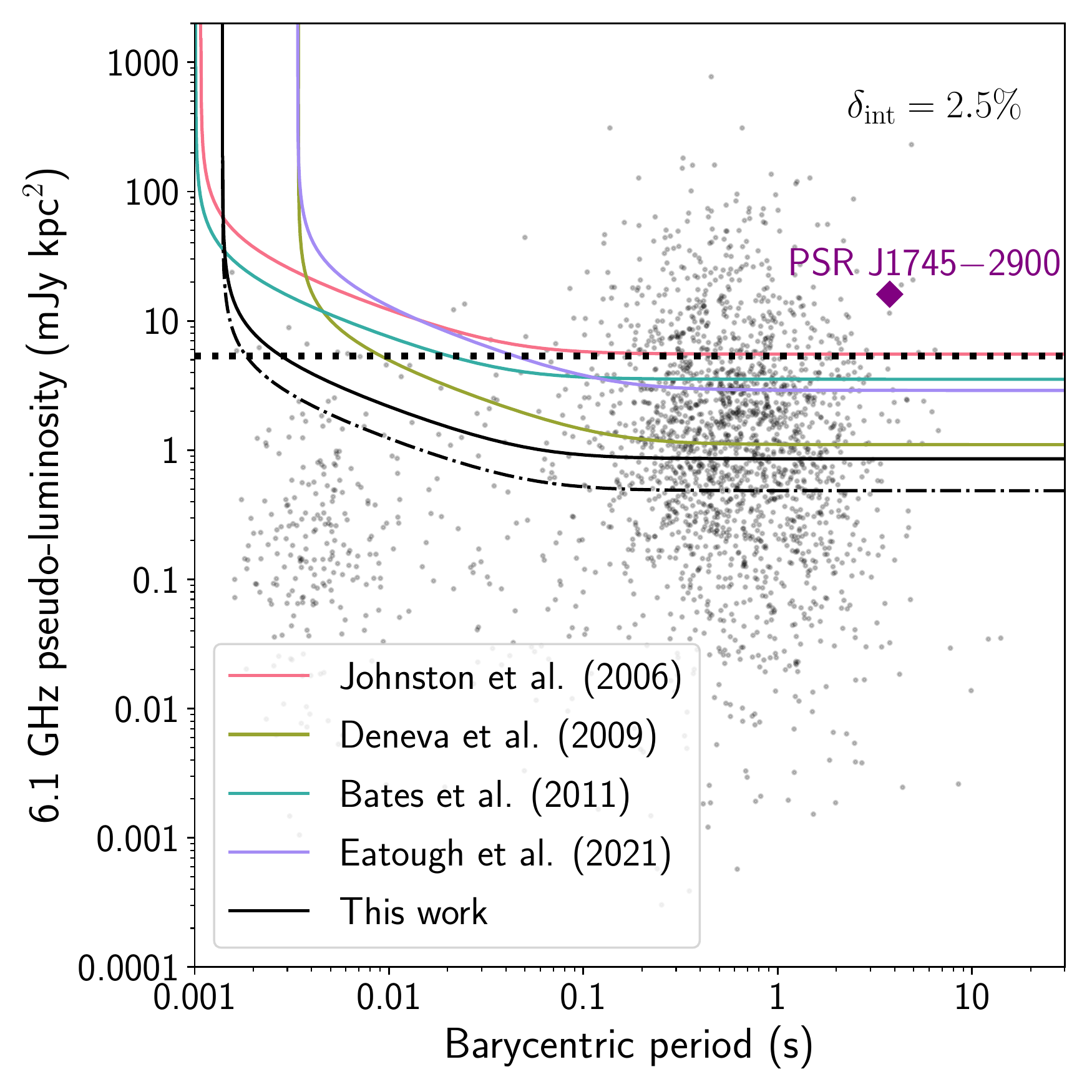}
\caption{$L_{\rm min}~(6.1~{\rm GHz})$ curves for different GC pulsar surveys (various colors). The solid curves represent $L_{\rm min}~(6.1~{\rm GHz})$ for GC pointings at the survey parameters listed in Table~\ref{fig4}. For \citet{Eatough2021}, the plotted curve corresponds to their 4.85~GHz observations. The dash-dot black curves indicates $L_{\rm min}$ for our $T=5$~minutes pointings away from the Sgr~A* complex. All curves assume $\delta_{\rm int} = 2.5\%$ (median duty cycle for CPs). The black dotted line marks the single harmonic sensitivity for our $T=30$~minutes GC scans. The background shows a scatter of 2197 pulsars with known 1.4~GHz flux densities ($S_{1.4}$) and known distances in the ATNF pulsar catalog v1.66. We scaled $S_{1.4}$ to 6.1~GHz assuming $S_{\nu} \propto \nu^{\alpha}$, where $\alpha = -1.4$ is the mean pulsar spectral index \citep{Bates2013}. The GC magnetar PSR~J1745$-$2900 is highlighted with a purple diamond marker. \label{fig7}}
\end{figure}
%%%%%%%%%%%%%%%%%%%%%%%%%%%%%%%%%
Applying the above power-law scaling and assuming $\delta_{\rm int} = 2.5\%$, Figure~\ref{fig7} shows $L_{\rm min}(6.1~{\rm GHz})$ for various surveys as function of $P_0$. Our survey clearly represents the most sensitive 4--8~GHz exploration for GC pulsars conducted to date. For $P_0 \gtrsim 100$~ms, our survey reaches down to $L_{\rm min} \approx 0.9$~mJy~kpc$^{2}$, i.e., a $18\%$ improvement over \citet{Deneva2009}, who also utilized the GBT for their observations. However, at $\nu_c = 4.8$~GHz, scattering limits the \citet{Deneva2009} sensitivity for $P_0 \lesssim 100$~ms, with $P_0 \leq 3.4$~ms pulsars rendered undetectable. In contrast, our survey still provides significant sensitivity to $P_0 \geq 1.35$~ms, thereby extending our discovery phase space to include potential superluminous MSPs ($L_{6.1} \geq 6$~mJy~kpc$^2$, $P_0 \approx 3$~ms) residing at the GC.

\subsection{Fake Pulsar Injection and Recovery}\label{sec:fakepsr}
Equations~\ref{eqn6} and \ref{eqn8} describe our survey sensitivity presuming ideal Gaussian, white noise backgrounds in dedispersed time series. However, real-world data frequently contains RFI and red noise, the latter of which substantially worsens our sensitivity to long $P_0$ \citep{Lazarus2015,vanHeerden2017}. Therefore, to measure our true survey sensitivity, we injected fake pulsars into our original dedispersed time series and verified their recovery. \\
%%% FIGURE 8
\begin{figure}[t!]
\includegraphics[width=0.48\textwidth]{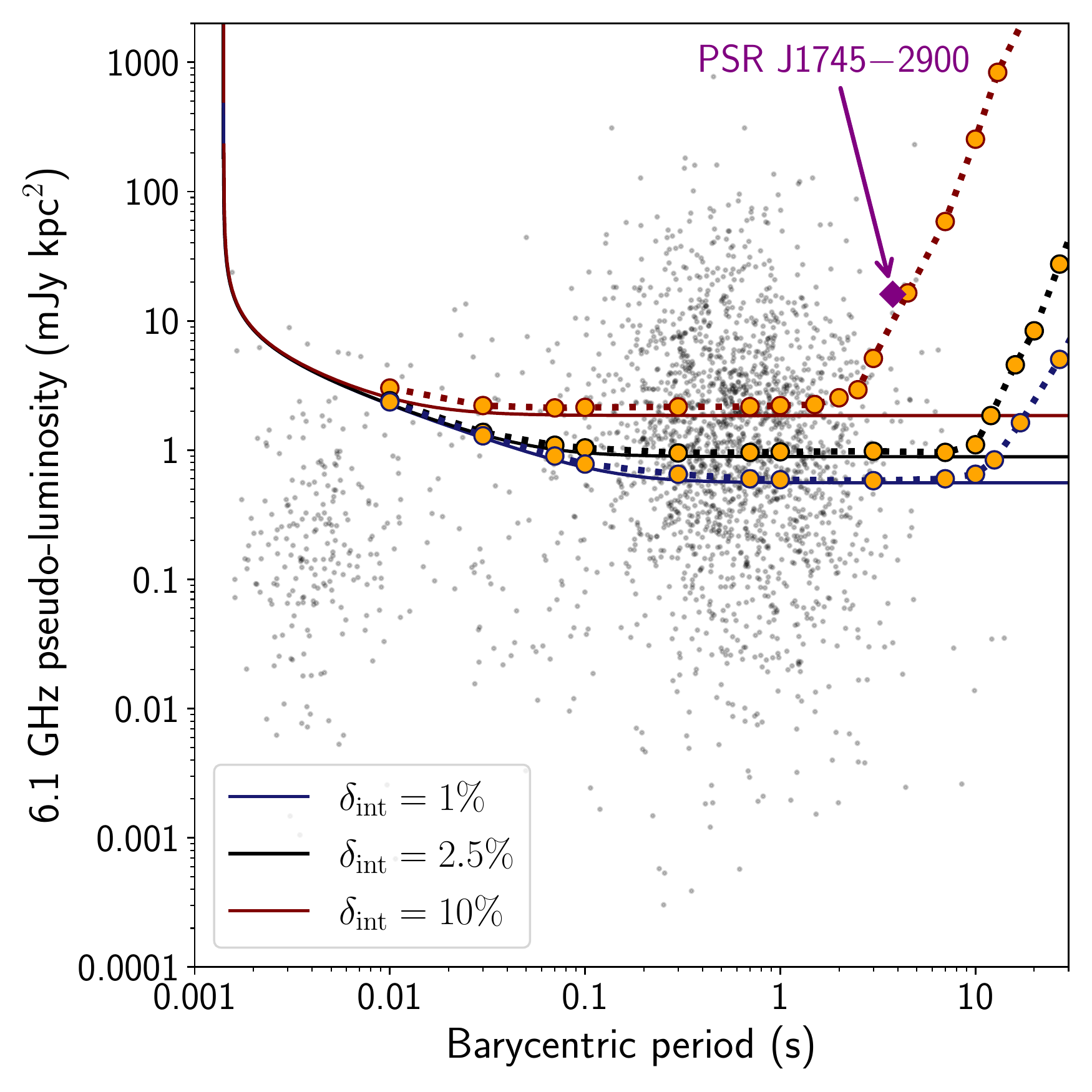}
\caption{Theoretical (solid curves) and true sensitivity estimates (dotted curves) for different $\delta_{\rm int}$ (various colors). All curves assume the survey parameters of our 30-minutes A00 scans mentioned in Table~\ref{tab4}. The duty cycles $\delta_{\rm int} = 1\%$ (dark blue) and $\delta_{\rm int} = 10\%$ (maroon) represent, respectively, the 10th and 93rd percentile of the empirical duty cycle distribution of CPs. The median $\delta_{\rm int}$ of the corresponding distribution is $2.5\%$ (black). The orange circular markers indicate minimum recovered 6.1~GHz pseudo-luminosities of fake pulsar injections at different trial $\delta_{\rm int}$ and $P_0$. As in Figure~\ref{fig7}, the background plot is a scatter of 2197 pulsars from the ATNF pulsar catalog v1.66. The GC magnetar PSR~J1745$-$2900 is labeled with a purple diamond marker. \label{fig8}}
\end{figure}
%%%%%%%%%%%%%%%%%%%%%%%%%%%%%%%%%

Following Section 2.2 of \citet{Suresh2021}, we first calibrated 11,016 dedispersed time series from six 30-minutes A00 scans at epoch~4 (MJD~58738). To simulate a fake pulsar, we began with an input $\delta_{\rm int}$, $P_0$, and a 6.1~GHz pseudo-luminosity $L_{6.1}$. We chose an initial pulse phase randomly from a standard uniform distribution. We then randomly drew Gaussian single pulse amplitudes $\mathcal{A}$ and FWHMs $W$ from lognormal distributions with means ($L_{6.1}/ \delta_{\rm int}d_{\rm GC}^2$) and $\delta_{\rm int}P_0$ respectively. For both distributions, we assumed standard deviations equal to $5\%$ of their respective means. To each $W$ sample, we then added $t_{\rm samp}$ in quadrature. Utilizing the drawn $\mathcal{A}$ and updated $W$ samples, we thus generated a periodic Gaussian pulse train that models a discretely sampled signal at the location of a pulsar (pulse jitter neglected). We next incorporated scattering by convolving the simulated Gaussian pulse train with a one-sided, decaying exponential of time scale, $\tau_{\rm sc}(6.1~{\rm GHz}) \simeq 1.35$~ms. Finally, we added the resulting periodic signal to all calibrated time series to complete our fake pulsar injection. In the above exercise, we ignored instrumental and dispersive pulse broadening as these effects are negligible for the parameters of our survey and data processing (see Appendix~\ref{appendix:Weff}). \\

For every trial $\delta_{\rm int}$ and $P_0$, we kicked off our pulsar injections at $L_{6.1} = 0.8L_{\rm min}$. Passing all calibrated time series containing the injected signal through our processing pipeline (including time series detrending), we stepped up $L_{6.1}$ in small increments until the detection significance of the artificial pulsar exceeded $6\sigma_{\rm ps}$. Let $L_{\rm min}^{\rm true}$ denote the minimum recovered $L_{6.1}$ for a given trial $\delta_{\rm int}$ and $P_0$. Averaging $L_{\rm min}^{\rm true}$ over identical fake pulsar injections in 11,016 calibrated time series, we obtained the dotted sensitivity curves shown in Figure~\ref{fig8}. \\

Noticeably, RFI raises our survey sensitivity above theoretical limits by 3--7$\%$ across all $P_0$. Moreover, the presence of red noise in our raw data progressively hinders our search sensitivity at longer $P_0$ and larger $\delta_{\rm int}$. However, despite the deleterious impact of red noise and RFI on our pulsar searches, our survey crucially retains adequate sensitivity to over $95\%$ of theoretically detectable pulsars for a median CP $\delta_{\rm int}$ of $2.5\%$.

% SECTION 5: SUMMARY AND DISCUSSION
\section{Summary and Discussion} \label{sec:disc}
We have conducted a comprehensive 4--8~GHz search of the central $6\farcm25$ ($\approx 14.9$~pc in projection) of our Galaxy for pulsars. Utilizing the GBT, our observing  program comprised of 11 $T\geq30$~minutes integrations on the GC and 53 $T=5$~minutes integrations on nearby Galactic bulge fields. As proof of our survey integrity, we successfully demonstrated the detection of the GC magnetar PSR~J1745$-$2900 in all of our GC scans. Executing Fourier-domain acceleration and jerk searches, we report the non-detection of hitherto unknown periodic astrophysical emissions in our data above a $6\sigma$ detection threshold.  \\

Our investigations constitute the most sensitive 4--8~GHz exploration for GC pulsars conducted to date. For $\delta_{\rm int} = 2.5\%$ and $P_0 \simeq 1$~s, our survey reaches down to $L_{\rm min}^{\rm  true} \approx 1$~mJy~kpc$^{2}$, i.e. a sensitivity improvement of at least $18\%$ over past GC pulsar searches conducted at similar radio frequencies. Notably, our observations open the window to discovering potential superluminous MSPs ($L_{6.1} \geq 6$~mJy~kpc$^{2}$, $P_0 \approx 3$~ms) at the GC. Though we focused our binary pulsar searches on CP orbits around Sgr~A* or stellar-mass BHs, our chosen processing parameters in Table~\ref{tab2} also incorporate searches for bright MSPs with low radial pulsar accelerations ($z_{\rm max} \lesssim 0.5 P_{0,{\rm ms}}^{-1}$) and jerks ($w_{\rm max} \lesssim 1.1 P_{0,{\rm ms}}^{-1}$). \\

Studying pulsar demographics in our Galaxy, \citet{Freire2013} argued that the globular cluster pulsar population is likely older than its Galactic field counterpart. Analogous to globular clusters, the GC environment, with its high density of $\sim 10^6$ stars per cubic parsec \citep{Schodel2018}, is predicted to favor MSP production. About $3\%$ of Galactic field MSPs have $L_{6.1} \geq L_{\rm min}^{\rm  true}$ of our survey.
Say that GC MSPs follow similar population-level statistics as their Galactic field equivalents. For a beaming fraction $f_b = 0.7$ \citep{Kramer1998}, our non-detection of GC pulsars therefore constrains the total MSP count in the central parsec of our Galaxy to $N_{\rm MSP} \lesssim 50$. Our $N_{\rm MSP}$ estimate is a factor of 20 smaller than that derived by \citet{Wharton2012} possibly due to our lack of sensitivity to tight MSP orbits. Sideband searches \citep{Ransom2003} and coherent full-orbit demodulation algorithms \citep{Allen2013,Balakrishnan2021} will both provide enhanced sensitivity to short binary orbital periods ($P_b \ll T$), thereby yielding stronger constraints on the GC MSP population in the near future. \\

Alternatively, our non-detection of GC pulsars can be attributed to complex pulsar orbital dynamics arising from plausible close encounters with other neutron stars, stellar-mass BHs, and intermediate-mass BHs in the dense GC environment. Such proximate compact object flybys can potentially disrupt binaries and scatter pulsars into hyperbolic orbits that lead away from the GC \citep{JialeLi2021}. On the other hand, a preference for magnetar formation \citep{Dexter2014} at the GC may explain the ``missing pulsar problem'' by virtue of the comparatively shorter magnetar lifetimes ($\sim 10^4$~years). \\

Throughout our study, we presumed that interstellar scattering does not limit GC pulsar surveys. While $\tau_{\rm sc}(\nu)$ measurements of the GC magnetar support the above premise, observations of pulse broadening along different lines of sight are necessary to build a more complete picture of the turbulent central ISM. Hence, we encourage regular monitoring of the GC for fast transients to obtain robust scattering constraints on the ionized central ISM, and thus better inform future GC pulsar surveys. 

%% IMPORTANT! The old "\acknowledgment" command is now depreciated. It was not robust enough to handle our new dual anonymous review requirements and thus been replaced with the acknowledgment environment. If you try to compile with \acknowledgment you will get an error print to the screen and in the compiled pdf.
\begin{acknowledgments}
A.S. thanks Scott M.~Ransom for timely responses to software-related queries. A.S., J.M.C., and S.C. acknowledge support from the National Science Foundation (AAG~1815242). J.M.C and S.C. are members of the NANOGrav Physics Frontiers
Center, which is supported by the NSF award PHY--1430284. Breakthrough Listen is managed by the Breakthrough Initiatives, sponsored by the Breakthrough Prize Foundation. The Green Bank Observatory is a facility of the National Science Foundation, operated under cooperative agreement by Associated Universities, Inc.\\ 
 
This work used the Extreme Science and Engineering Discovery Environment (XSEDE) through allocations PHY200054 and PHY210038, which is supported by National Science Foundation grant number ACI$-$1548562. Specifically, it used the Bridges-2 system, which is supported by NSF award number ACI$-$1928147, at the Pittsburgh Supercomputing Center (PSC).
\end{acknowledgments}

%% Following the acknowledgments section, use the following syntax and the\facility{} or \facilities{} macros to list the keywords of facilities used in the research for the paper.  Each keyword is check against the master list during copy editing.  Individual instruments can be provided in  parentheses, after the keyword, but they are not verified.

\vspace{5mm}
\facilities{GBT, XSEDE \citep{XSEDE}.}

\software{Astropy \citep{Astropy2013, Astropy2018},
          NumPy \citep{NumPy},
          Matplotlib \citep{Matplotlib},
          {\tt PRESTO} \citep{PRESTO},
          Python~3 (\url{https://www.python.org}), 
          SciPy \citep{SciPy}.
         }

% APPENDIX 
\appendix
\restartappendixnumbering
\section{Pulse-Averaged Profiles of Sample Periodicity Candidates}\label{appendix:avg_profiles}
\begin{figure*}[ht!]
\figurenum{A1}
\includegraphics[width=\textwidth]{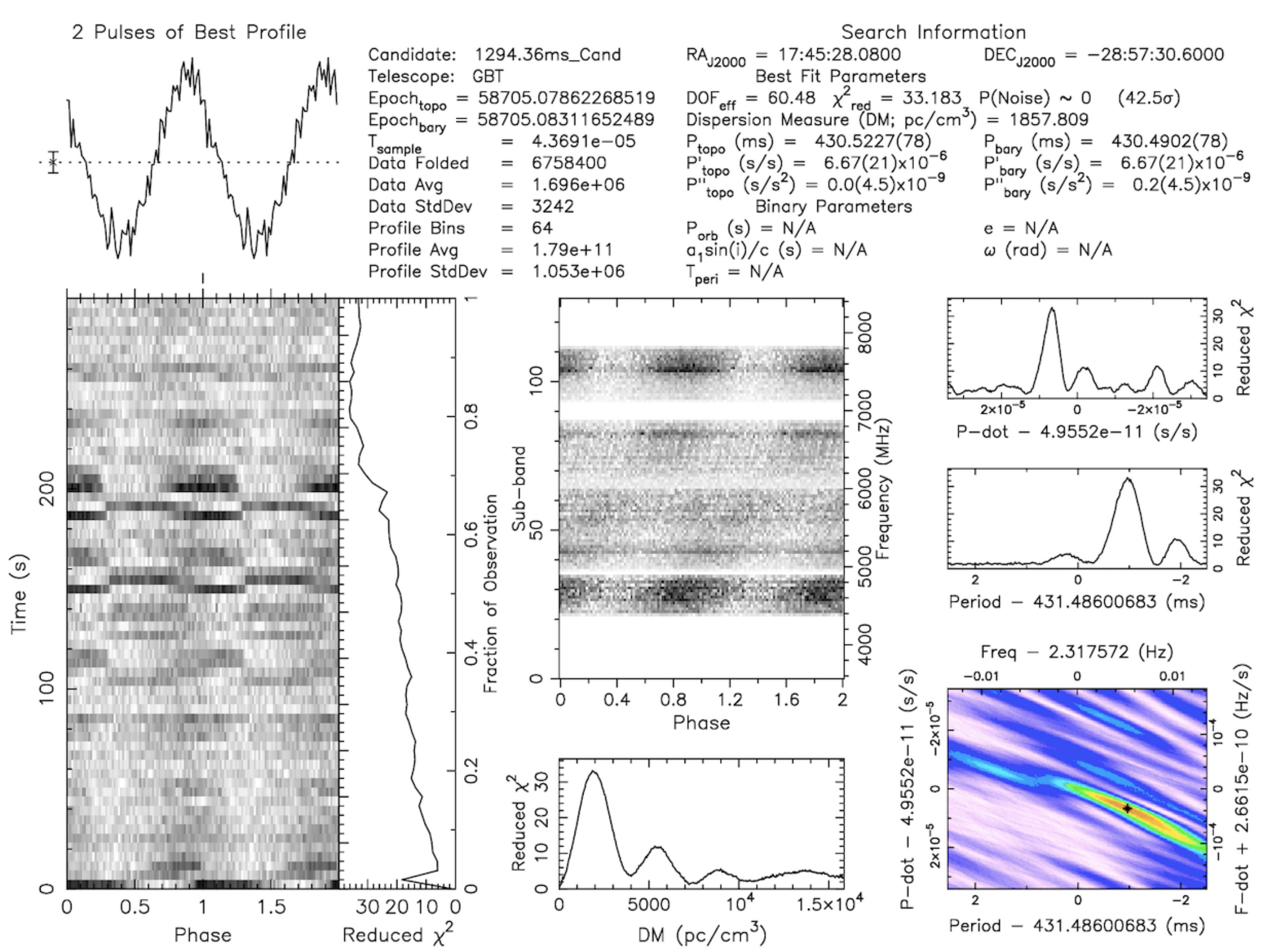}
\caption{A sample periodicity detection in pointing C04 that falls within the sea of candidates between $P=2$~ms and $P=100$~ms in the $P$--DM plane (see Figure~\ref{fig5}). Top left panel: Average pulse profile of candidate. Bottom left panel: Rotation-resolved profile with flux density shown on the grayscale. The reduced $\chi^2$ measures the departure of the pulse-averaged profile from a flat noisy model. Top middle panel: Phase-resolved dynamic spectrum of candidate. Bottom middle panel: Variation of the reduced $\chi^2$ with DM. Top right panel: Reduced $\chi^2$ vs. period derivative ($\dot{P}$). Central right panel: Reduced $\chi^2$ vs. folding period ($P$).  Bottom right panel: Raster plot of the reduced $\chi^2$ in the $P$--$\dot{P}$ plane. The pulsation significance (42.5$\sigma$) quoted in the top right corner quantifies the equivalent Gaussian detection probability determined from the reduced $\chi^2$ of the average profile. \label{figA1}}
\end{figure*}

\begin{figure*}[ht!]
\figurenum{A2}
\includegraphics[width=\textwidth]{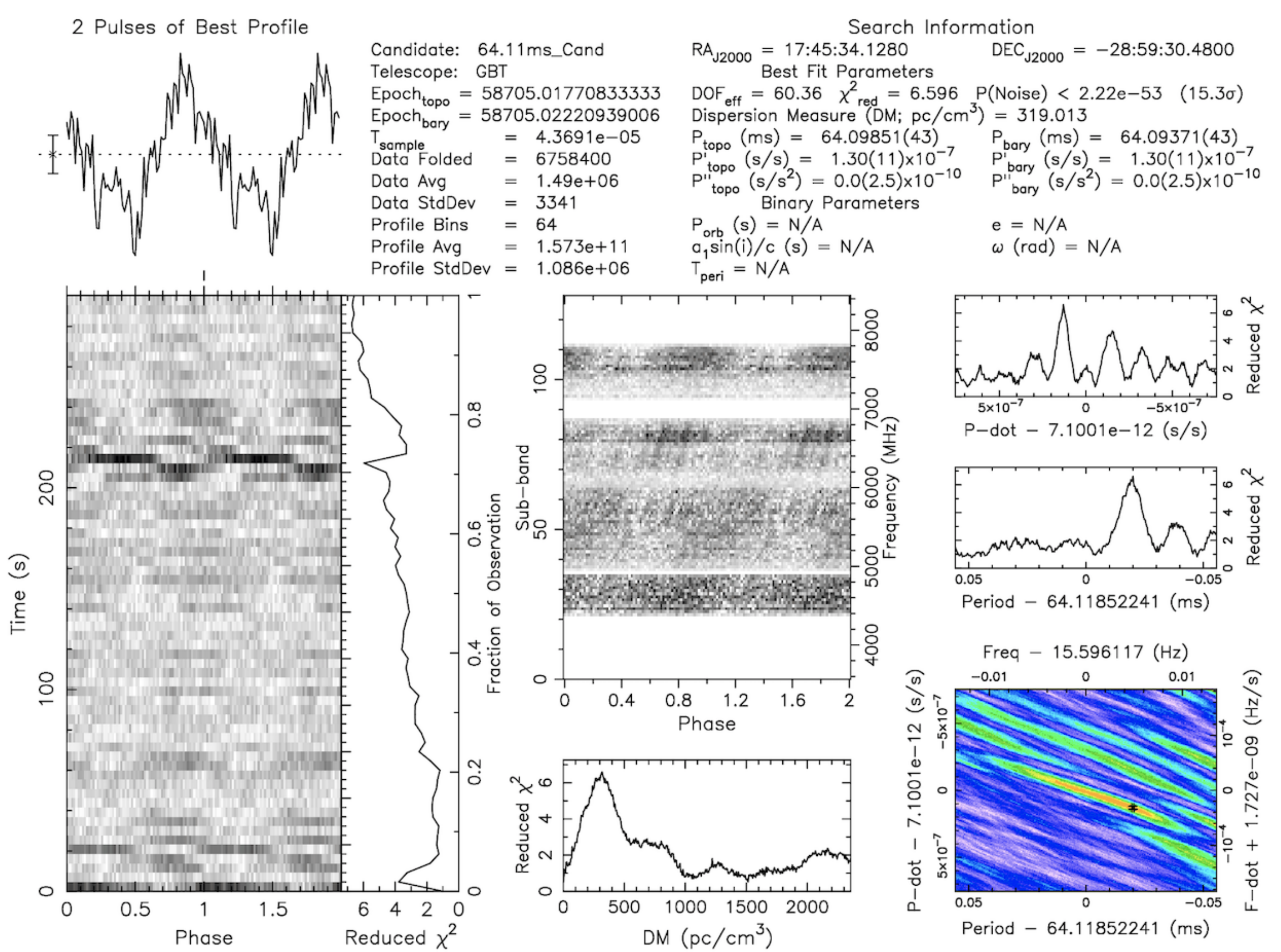}
\caption{A {\tt prepfold} output of a periodicity detection in pointing B02. \label{figA2}}
\end{figure*}

Figures~\ref{figA1} and \ref{figA2} show average profiles of two sample candidates that sit amidst the large scatter of detections between $P=2$~ms and $P=100$~ms in the $P$--DM plane of Figure~\ref{fig5}. Average profiles were generated using the {\tt prepfold} routine of {\tt PRESTO} \citep{PRESTO}.
%%% FIGURE A1
%%%%%%%%%%%%%%%%%%%%%%%%%%%%%%%%%% 
%%% FIGURE A1
%%%%%%%%%%%%%%%%%%%%%%%%%%%%%%%%%% 

\section{Effective Single Pulse Widths in Dedispersed Time Series}\label{appendix:Weff}
Consider a pulsar emitting single pulses of average intrinsic width $W_{\rm int}$. Accounting for instrumental effects, sampling, and propagation-induced broadening, the effective pulse width in a dedispersed time series is
%%% EQUATION A1
\begin{align}\label{eqnA1}
W_{\rm eff} = \left(W_{\rm int}^2 + t_{\rm samp}^2 + t_{R}^2 + t_{\rm chan}^2 + t_{\rm BW}^2 + \tau_{\rm sc}^2 \right)^{1/2}.  
\end{align}
%%%%%%%%%%%%%%%%%%%%%%%%%%%%%%%%%
Here, $t_{\rm samp}$ is the sample interval in the dedispersed time series, and $t_R \sim (\Delta \nu_{\rm ch})^{-1}$ is the receiver filter response time for a channel bandwidth $\Delta \nu_{\rm ch}$. \\

The term $t_ {\rm chan}$ represents the intrachannel dispersive smearing given by \citep{Cordes2003}
%%% EQUATION A1
\begin{align}\label{eqnA2}
t_{\rm chan} \approx 8.3~\mu{\rm s} \left(\frac{{\rm DM}_{\rm pc~cm^{-3}} \ \Delta \nu_{\rm ch, MHz}}{\nu_{c,\rm GHz}^3} \right).
\end{align}  
%%%%%%%%%%%%%%%%%%%%%%%%%%%%%%%%%% 
Further, $t_ {\rm BW}$ quantifies the residual broadband dispersive delay for a DM error $\delta \rm DM$.
%%% EQUATION A3
\begin{align}\label{eqnA3}
t_{\rm BW} \approx 8.3~\mu{\rm s} \left(\frac{{\rm \delta DM}_{\rm pc~cm^{-3}} \ B_{\rm MHz}}{\nu_{c,\rm GHz}^3} \right).
\end{align}   
%%%%%%%%%%%%%%%%%%%%%%%%%%%%%%%%%% 
Finally, $\tau_{\rm sc}(\nu)$ is the pulse-broadening time scale from multi-path wave propagation through turbulent ionized plasma. \\

For our observations, $\nu_c \approx 6.1$~GHz, $B\approx 3.35$~GHz, $\Delta \nu_{\rm ch} \approx 91.67$~kHz, and $t_{\rm samp} \approx 349.53~\mu$s. Working with a DM step size of 3~pc~cm$^{-3}$, our trial DMs are at best, $\delta {\rm DM} = 1.5$~pc~cm$^{-3}$ off from the true DM of a source. Collectively, the above numbers imply $t_R \sim 11~\mu$s, $t_{\rm BW} \approx 184~\mu$s, and $t_{\rm chan} \approx 6~\mu$s for ${\rm DM} = 1776$~pc~cm$^{-3}$ of the GC magnetar \citep{Suresh2021}. Assuming the empirical GC magnetar scattering law \citep{Spitler2014} given in Equation~\ref{eqn10}, $\tau_{\rm sc} \simeq 1.35$~ms at 6.1~GHz. \\

Since $\tau_{\rm sc} \gg t_{\rm chan},~ t_{\rm BW},~t_R$, Equation~\ref{eqnA1} can thus be simplified to
%%% EQUATION A4
\begin{align}\label{eqnA4}
W_{\rm eff}(\nu) = \left(W_{\rm int}^2 + t_{\rm samp}^2 + \tau_{\rm sc}^2(\nu) \right)^{1/2}.
\end{align}
%%%%%%%%%%%%%%%%%%%%%%%%%%%%%%%%%

\bibliography{references}
\bibliographystyle{aasjournal}

\end{document}